\documentclass[a4paper, 10pt, conference]{ieeeconf}      % Use this line for a4
\IEEEoverridecommandlockouts                              % This command is only
\usepackage[utf8]{inputenc}
\usepackage[T1]{fontenc}
\usepackage{graphicx}
\usepackage{authblk}
\title{\LARGE \bf
Impact on the Productivity of Remotely Working IT Professionals of Bangladesh during the Coronavirus Disease 2019
}

\author{Kishan Kumar Ganguly,
Noshin Tahsin, Mridha Md. Nafis Fuad, Toukir Ahammed, \authorcr Moumita Asad, Syed Fatiul Huq, A.T.M. Fazlay Rabbi and
Kazi Sakib}
\affil{Institute of Information Technology,
University of Dhaka,
Dhaka, Bangladesh \authorcr Email:  \tt \{kkganguly, bsse0914, bsse0920, bsse0806,\authorcr \tt bsse0731, bsse0732, bsse0926, sakib\}@iit.du.ac.bd\vspace{0.1cm}}

\begin{document}

\maketitle
\thispagestyle{empty}
\pagestyle{empty}

%%%%%%%%%%%%%%%%%%%%%%%%%%%%%%%%%%%%%%%%%%%%%%%%%%%%%%%%%%%%%%%%%%%%%%%%%%%%%%%%
\begin{abstract}
Similar to the rest of the world, the recent pandemic situation has forced the IT professionals of Bangladesh to adopt remote work. The aim of this study is to find out whether remote work can be continued even after the lockdown is lifted. As work from home may change various productivity related aspects of the employees, i.e., team dynamics and company dynamics, it is necessary to understand the nature of the change during WFH. Conducting a survey, we asked the IT professionals of Bangladesh how they perceive their level of productivity during WFH and how the factors related to productivity have changed. We analyzed the change and identified the areas affected by WFH. We discovered that resource and workspace related issues, emotional well-being of the employees have been hampered the most during WFH. We believe that the findings from this study will help to decide how to resolve those issues and will help to understand whether WFH can be continued even after the lockdown is lifted.

\end{abstract}

%%%%%%%%%%%%%%%%%%%%%%%%%%%%%%%%%%%%%%%%%%%%%%%%%%%%%%%%%%%%%%%%%%%%%%%%%%%%%%%%
\section{INTRODUCTION}
%by directly and indirectly impacting the market in the form of %software/ hardware products and IT enabled services.

With a GDP growth rate of 8.2\%, Bangladesh has one of the fastest growing market economies in the world \cite{gdpWB}. Current Digital Economy Report by the United Nations show that the ICT sector has become a major economic driver as global ICT deliverable service exports have tripled from 2005 to 2018 \cite{derUN}. The export value of Bangladesh ICT sector has correspondingly increased about fivefold in the same time range \cite{gdpBD}. Moreover, the ICT sector employment in Bangladesh has increased three times starting from 2013 and still growing \cite{itesBD}. The target of \$5 billion ICT export and 2 million employment set forth by Bangladesh indicates that its ICT sector will continue to grow.

Due to the current coronavirus pandemic, ICT companies worldwide are increasingly adopting Work From Home (WFH). A recent survey on global technology industry employees has shown that WFH during the pandemic has exercised both positive effects such as increased short breaks and more freedom as well as negative ones such as burnout, resource problems and mental health issues \cite{remST}. However, these factors vary across countries, for example, countries with more nuclear families may face less distraction related productivity loss than the ones with joint families. A few studies have been conducted regarding the influence of these factors on productivity which only focus on developed countries in a pandemic setting \cite{ralph2020pandemic}. To the best of the author's knowledge, no work has been done on this subject for developing countries like Bangladesh to date. However, such a study on Bangladesh is required due to the following reasons. Firstly, the rapid growth of the ICT sector in Bangladesh has created increasing pressure on the the ICT companies to be more productive even during the pandemic. As these companies are increasingly adopting WFH, knowing the factors that control productivity can help them to optimize employee performance. Secondly, Bangladesh is a densely populated country with a significantly different culture from the developed countries studied in the literature. Therefore, the influence of the socio-cultural environment on productivity related factors will be evident in a study conducted on Bangladesh. Thirdly, As Bangladesh has been strongly hit by this pandemic, the employees are burdened with increased mental pressure.

The study conducted by Ralph et al. is the only one that analyzes ICT sector employee productivity during COVID- 19 \cite{ralph2020pandemic}. They measured the impact of mental wellbeing during this pandemic on employee productivity and found a significantly negative relationship. Apart from this, several studies were conducted on productivity factor analysis before the pandemic. Wagner et al. performed a systematic review of the productivity related factors where they divided these into two groups namely technical and soft factors \cite{wagner2018systematic}. Technical factors are related to the project characteristics and developing process such as software size and project duration \cite{wagner2018systematic}. Soft factors comprise the culture and working environment related ones such as communication and cooperation between team members \cite{wagner2018systematic}. Graziotin et al. assessed the productivity of software engineers based on their comments on news and social media \cite{graziotin2014software}. They found that increasing task and turnover rate decreases productivity where frequent short breaks increases it.  Murphy et al. analyzed the productivity of employees from Google, ABB and National Instruments \cite{murphy2019predicts}. They observed three factors to be positively associated with productivity which are meeting preparation, feedback and time management freedom. Meyer et al. conducted a survey on 379 participants from IT companies of developed countries where they found the role of workplace environment to significantly influence employee productivity \cite{meyer2014software}. As literature shows, no study has been done on the productivity ICT professionals of a developing country during the pandemic situation. 

The objective of this study is to understand the impact of COVID-19 pandemic on the productivity. The target population is the remotely working IT professionals of Bangladesh. First, an anonymous survey is designed to collect the data. The questionnaire is prepared considering the factors that have an impact on productivity in literature. For example, miscommunication has negative impact on productivity, quiet workplace has positive impact on productivity etc. A pilot survey is conducted to refine the questionnaire. Experts from IT industry and academicians from universities were the participants of this pilot survey. Then the final questionnaire is circulated among the IT professionals of Bangladesh. The data is collected throughout a month. After collecting data, the preprocessing, i.e., data cleaning and encoding, is performed. Finally, descriptive analysis is conducted on each of the factors collected from the survey.

In this survey, the respondents were asked 47 questions about their personal-life and work-life to discover how those aspects affect productivity during WFH. Among those, 24 questions were related to their work-life, which were carefully chosen based on related studies mentioned in the previous paragraph. After analyzing those 24 aspects, it has been found that WFH can be adopted in the software industry of Bangladesh as it increases the overall employee productivity. In this study, the top 6 influential factors that positively impact productivity during WFH has been carefully examined to find whether there is a decrease in productivity caused by those factors. For example, team cooperation is one of those factors.  After analyzing the survey responses, it has been found that 60.7\% of the respondents’ team cooperation either increased or remained unchanged during WFH. As a result, productivity is not hampered.  Similarly, the other 5  factors increased or remain unchanged, not hampering employee productivity. To determine whether WFH increases productivity or not, it is necessary to examine the factors negatively affecting productivity. Related studies show that miscommunication is such a factor. From the data, it has been seen that although there is an increase in miscommunication among the employees, the top 6 factors positively affecting productivity have increased. As a result,  increased miscommunication could not hamper productivity. Other factors that negatively affect productivity have been examined also. It is seen that resource accessibility, work environment, and emotional well-being during WFH have changed the most leaving a negative impact on productivity. Most of the respondents are facing worsened internet connections (60.8\%), hardware constraints (60.2\%), more interruption (49.4\%), the necessity of a quiet work-space (42.6\%) and increased frequency of household chores (65.2\%). Most of them are feeling less cheerful (57.7\%) than before. As a result, their productivity has been decreased. From all these findings, it can be seen that WFH does not affect 18 out of 24 of the productivity-related factors.  The data reflects that the major problems during WFH are resource-related, work-environment related and mental state-related. The study reveals that ensuring strong internet connection, suitable workspace and emotional well-being, the drawbacks of WFH can be mitigated. 

\section{Methodology}
After problem identification and thorough background study, a questionnaire was constructed. After incorporating the feedbacks of a pilot survey, the final survey was conducted. After the pre-processing of the collected data, patterns and relationship between the factors have been discovered through descriptive analysis. The steps are described in the following sub-sections.

\begin{enumerate}
    \item 
    \textbf{Questionnaire construction}: After thoroughly studying related works, 24 factors related to employee productivity were identified. Questions and answers were designed to understand the change in each of those factors during WFH and relate that to productivity. The questions were asked in such a way that makes it easier for the respondents to compare how the factors were before WFH and how the factors are now. To scale the responses, the Likert scale, the most widely-used approach has been chosen for this study. For example, to understand how the feedback of work has changed, the respondents were asked “During work from home, I get useful feedback about job performance.” and they were told to select one of the following options: a) Less b) Slightly less c) Same as before d) Slightly more e) More.

\item
\textbf{Pilot survey}: To refine the questionnaire, a pilot survey was conducted. Experts from the IT industry and academicians from universities participated in this pilot survey and evaluated the questionnaire. Incorporating the valuable feedback received from them, the questionnaire was modified accordingly. 

\item
\textbf{Survey conduction}: After the refinement, the final questionnaire was circulated among the IT professionals of Bangladesh. To keep the dataset unbiased, data was collected from IT professionals across the country. The respondents were from 29 out of 64 districts of Bangladesh. The respondents were from diverse company sectors including software firms, banks and other financial organizations, telecom companies, technology consultancies, IT and network services providers, and Information Technology Enabled Services (ITeS) \cite{ami2020effects}. Responses were collected throughout a month. 

\item
\textbf{Preprocessing}: After collecting responses throughout a month, an excel datasheet was generated from the survey data. The initial column names were replaced with short and appropriate terms. The unnecessary columns, i.e., the email of the respondents and their company name were dropped for anonymization. As the study is concerned only with the professionals who are working from home during this pandemic, the rows containing data of the respondents who said they were not working remotely (106 respondents) and some invalid responses of people who were not IT professional (69 responses), were removed from the dataset. For this reason, this study is based on 1062 respondents who are working remotely, among the randomly sampled 1237 total respondents. There were some missing values in some columns. In that case, the analysis was based on the available responses. The column values have been encoded as most of the data was obtained from user selection from specified options, thus categorical. The answers were encoded according to Likert-scale as the options of the questions were designed based on it. The designation column was converted to a numeric category by manual verification as the respondents were free to set the designation title according to their workplace hierarchy. 

\item
\textbf{Analysis}: After pre-processing the data, descriptive analysis has been conducted on each of the factors collected from the survey to get an idea of the data and discover patterns and relationships among the factors. Pie charts, bar-charts, and stacked-bar charts have been developed to visualize the data.

\end{enumerate}

\section{Result and Discussion}
 The detailed results in the following sub-sections demonstrate the viability of WFH in the long run. Although WFH has both positive and negative consequences, necessary steps can be undertaken to negate the unwanted outcomes. The chosen 24 aspects related to the respondents’ work-life were categorized into 8 different sections: Team Dynamics, Company Dynamics, Team Collaboration, Access to Resources, Work Environment, Emotional Well-being, Proximity to the COVID-19 virus, and others, based on the relation between the factors to make the analysis more efficient. The dataset has been carefully examined to find which of the 8 sections are negatively affected by WFH so that it may help the authority to take necessary measures to mitigate the negative impacts and ensure maximum productivity during WFH. From the graphs, it has been found that team cooperation, activity level, well-defined goals, the number of task assignments, feedback of work, and frequency of communication are the most influential 6 factors positively impacting productivity. The changes in these factors have also been examined carefully. After analyzing the data, it has been found that the respondents’ resource accessibility, work environment, and  emotional well-being during WFH have changed the most leaving a negative impact on productivity. The data indicates that ensuring necessary resources, suitable work environment, and emotional well-being, the drawbacks of WFH can be mitigated. The following 8 sub-sections state the findings from the data with necessary graphs.

\begin{enumerate}

\item \textbf{Team Dynamics}:
 As related studies show, change in team dynamics plays a significant role in the distortion of individual member output \cite{wagner2018systematic}, it is necessary to track its change during WFH. To understand the nature of team dynamics during WFH, the respondents were asked about the change in their team cooperation, the definition of team goals, and team size. Fig. 1 is the visual representation of the change in team dynamics. During WFH, changes in the 3 factors related to team dynamics are stated below.

\begin{enumerate}
\item  \textbf{Team Cooperation}:
 Related studies show that, if cooperation between team members increases, productivity increases \cite{wagner2018systematic}.
 From the data, it is seen that 60.7\% of the respondents’ team cooperation either remained unchanged or increased during WFH. It can be said that there has been no major negative change in productivity due to this factor.

\item  \textbf{Definition of Team Goals}:
 According to related studies, a team with clearly-defined goals shows better performance and is more productive \cite{wagner2018systematic}. From the data, it is seen that 71.8\%  of the respondents’ goals are either equally clearly defined or more clearly defined during WFH. There has been no major negative change in productivity due to change in the teams' goal definition.

\item  \textbf{Team Size }:
According to literature, small teams are more efficient as increase in team size hampers communication and decreases productivity \cite{wagner2018systematic}. From the data, it is seen that 76\%  of the respondents’ team size is the same as before or has decreased during WFH which has a positive impact on productivity. 
\end{enumerate}

\item  \textbf{Company Dynamics}:
Related studies show that, changes in company dynamics such as a change in salary, number of task assignment, and turnover rate can affect an employee’s productivity to a great extent \cite{graziotin2014software}. To understand the nature of company dynamics during WFH, the respondents were asked about the change in their salary, the number of task assignments, and the turnover rate. Fig. 2 is the visual representation of the change in company dynamics. During WFH, changes in the 3 factors related to company dynamics are stated below.

\begin{enumerate}
\item  \textbf{Salary Change}:
According to literature, increasing salary increases productivity \cite{ouimet2020wages}. From the data, it can be seen that 78.9\% of the respondents’ salary is the same as before or has increased during WFH which has no negative effect on their productivity. 

\item  \textbf{Number of Task Assignment}:
Related studies show if employees are assigned more tasks than before, their productivity decreases \cite{graziotin2014software}. But from our data, it is seen that the respondents’ productivity increased with increasing number of tasks. It happened because the respondents perceived the term "Number of assigned tasks" as the number of tasks accomplished by them and thought it as a way to be more productive. 79.1\% of the respondents said that the number of assigned tasks has increased or is the same as before during WFH increasing their productivity. That is 79.1\% of the respondents accomplished more or the same amount of task during WFH.

\item  \textbf{Turnover Rate}:
While according to the literature, increase in the turnover rate decreases productivity \cite{graziotin2014software}, 88\% of the respondents said that the turnover rate in their company is the same as before or has decreased which has a positive impact on productivity.

\end{enumerate}

\begin{figure}[htp]
    \centering
    \includegraphics[width=8cm]{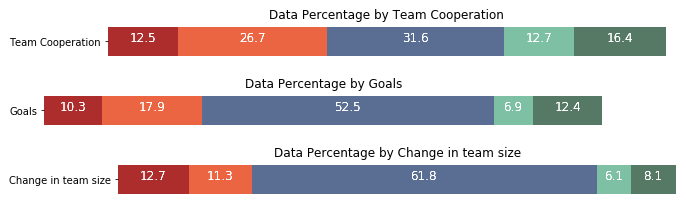}
    \caption{The change in team dynamics during WFH}
    \label{fig:galaxy1}
\end{figure}

\begin{figure}[htp]
    \centering
    \includegraphics[width=8cm]{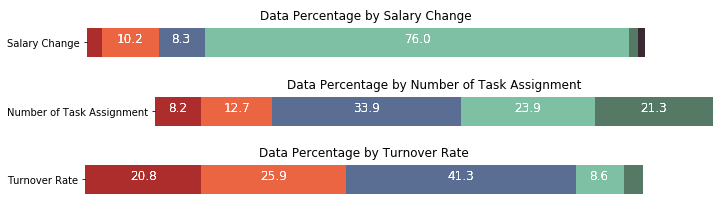}
    \caption{The change in company dynamics during WFH}
    \label{fig:galaxy2}
\end{figure}

\item \textbf{Team Collaboration}: To understand how well the team members are collaborating and communicating during WFH, the respondents were asked about the change in their frequency of miscommunication, frequency of communication, meeting preparation, and feedback of work, as literature shows that these factors are related to employee productivity \cite{graziotin2014software} \cite{murphy2019predicts} Fig. 3 is the visual representation of the change in team collaboration. During WFH, changes in the 4 factors related to team collaboration are stated below.

\begin{enumerate}
\item \textbf{Frequency of Miscommunication}:
Related studies show that increasing miscommunication decreases productivity \cite{graziotin2014software}. The data reveals that 60\% of the respondents’ frequency of miscommunication either remained unchanged or decreased during WFH. 40\% of the respondents said that there has been an increase in miscommunication. However, the most influential 6 factors positively affecting productivity have increased minimizing the negative impact of miscommunication on productivity.

\item \textbf{Frequency of communication}:
According to the literature, increase in communication between team members increases productivity \cite{wagner2018systematic}. From the data, it is seen that 60.6\% of the respondents’ frequency of communication has either increased or  remained unchanged leaving no negative impact on productivity.

\item
\textbf{Meeting Preparation}:
Taking more preparation for meetings increases productivity, according to related studies \cite{murphy2019predicts}. From the data, it is seen that 73.4\% of the respondents are taking the same or more preparation for meetings during WFH. As a result, there is no decrease in productivity due to the lack of proper meeting preparation. 

\item\textbf{Feedback}:
Literature shows that receiving useful feedback about job performance increases productivity \cite{murphy2019predicts}. From the data, it is seen that 67.5\% of the respondents are receiving the same or more useful feedback for their work during WFH. There is no decrease in productivity due to the lack of useful feedback.

\end{enumerate}

\item
\textbf{Access to Resources}: Whether the employees have access to necessary resources, is important to be considered when measuring productivity. For understanding how well the respondents are getting access to resources during WFH, the respondents were asked about their software accessibility, quality of internet connection, and hardware accessibility. Fig. 4 is the visual representation of the change in access to resources. During WFH, changes in the 3 factors related to resource access are stated below.

\begin{enumerate}

\item
\textbf{Software Accessibility}:
Related studies show, access to a limited number of tools (software) has a negative impact on productivity \cite{maxwell2000benchmarking}. Whether the respondents can access all the necessary software tools during WFH, is a major concern to understand the change in productivity. The data reveals that, 77.7\% of the respondents are not feeling any lack of software access during WFH.

 \item 
   \textbf{Quality of Internet Connection}: A weak internet connection hampers employee productivity to a great extent. From the survey data, a significant negative change has been noticed in the quality of internet connection during WFH. 60.8\% of the respondents said that the quality of internet connection of their home is worse than office causing an obvious decrease in productivity.

\item
\textbf{Hardware Accessibility }:
Hardware constraints negatively affect productivity. In this survey, 60.2\% of the respondents said they are facing hardware constraints during WFH. Among them 6.6\% are facing it always, 14.7\% are facing it often and 38.9\% are facing it sometimes. 

\end{enumerate}

\begin{figure}[htp]
    \centering
    \includegraphics[width=8cm]{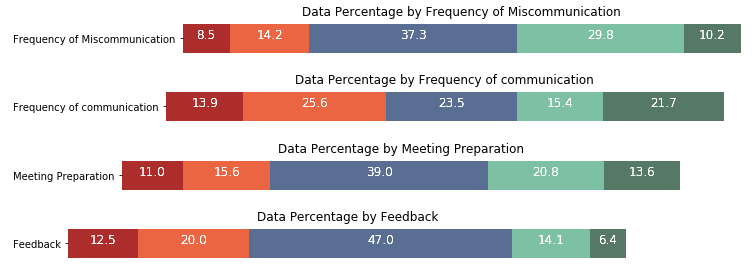}
    \caption{The change in team collaboration during WFH}
    \label{fig:galaxy3}
\end{figure}

\begin{figure}[htp]
    \centering
    \includegraphics[width=8cm]{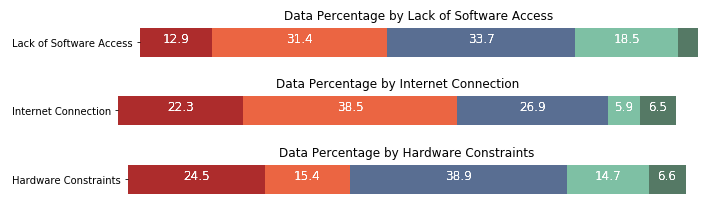}
    \caption{The change in access to resources during WFH}
    \label{fig:galaxy4}
\end{figure}

\item
\textbf{Work Environment}:
A proper work environment is vital for the employees for working productively. Studies show that the workplace needs to be quiet, and free of distractions \cite{wagner2018systematic}. The lack of a proper work environment hampers productivity. To find how work environment during WFH differs from the office environment, and how it is impacting productivity, the respondents were asked about their workplace suitability and the frequency of interruptions they are facing. Fig. 5 is the visual representation of the change in work environment. During WFH, changes in the 2 factors related to  the work environment are stated below.

\begin{enumerate}
    \item 
   \textbf{Workplace Suitability}:
Literature shows a quiet work environment increases productivity \cite{wagner2018systematic}. From the data, it is seen that 42.6\% of the respondents are feeling the necessity of a quite work-space. This portion of the respondents’ productivity has been hampered. 32.1\% of the respondents said that the workplace is more quite than the office.  25.3\% said it is suitable during WFH as it was before.

\item
\textbf{Frequency of Interruptions}:
According to related studies, the less the frequency of interruptions, the more productivity is \cite{meyer2014software}. From the data, it is seen that, 49.4\% of the respondents are feeling that they are facing more interruption during WFH, while 27.1\% said they are facing less interruption. This increase in interruptions caused decrease in their productivity.

\end{enumerate}

\item
\textbf{Emotional Well-being}:
Studies show that the emotional well-being of an employee is associated with productivity \cite{topp20155}. If a person is emotionally stable and cheerful, he can perform his tasks better than an emotionally drained person. To understand the state of the emotional well-being of the employees during WFH, the respondents were asked whether they are feeling cheerful, calm, active, whether they are waking up feeling rested every day, and whether they are feeling interest in their daily activities. Fig. 6 is the visual representation of the change in the employees' emotional well-being. During WFH, changes in the 5 factors related to emotional well-being are stated below.

\begin{enumerate}
    \item 
\textbf{Cheerfulness}:
Being cheerful and in good spirits increases productivity. 57.7\% of the respondents said that they are feeling less cheerful than before, which has a negative impact on productivity during WFH. 21.7\% of the respondents said that they are feeling more cheerful, while the rest of the respondents said their cheerfulness is the same as before.

\item
\textbf{Calmness}:
According to related studies, feeling calm and relaxed has a positive impact on productivity. From the data, it is seen that 51\% of the respondents said that they are feeling calm and relaxed, either the same as before or more than before. As a result, there is no negative impact on their productivity.

\item
\textbf{Activity level}: Related studies show that, feeling active and vigorous increases productivity. From the data, it is seen that 53.5\%
of the respondents are feeling active and vigorous, either the same as before or more than before. As a result, their productivity is not hampered.

\item
\textbf{Waking up rested}:
Related studies show, waking up feeling fresh and rested increases productivity. 62.3\% of the respondents said that they are waking up feeling more rested or the same as before, which has a positive impact on productivity.  

 \item
\textbf{Pursuance of interest}:
Respondents were asked whether their daily life is filled with things that interest them, as related studies show that pursuance of interest increases productivity. From the data, it is seen that 62.2\% of the respondents either agree or stay neutral of the fact that they are doing things that interest them, while 37.8\% of the respondents disagree.

\end{enumerate}

\begin{figure}[htp]
    \centering
    \includegraphics[width=8cm]{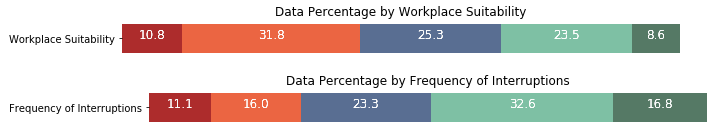}
    \caption{The change in work environment during WFH}
    \label{fig:galaxy7}
\end{figure}

\begin{figure}[htp]
    \centering
    \includegraphics[width=8cm]{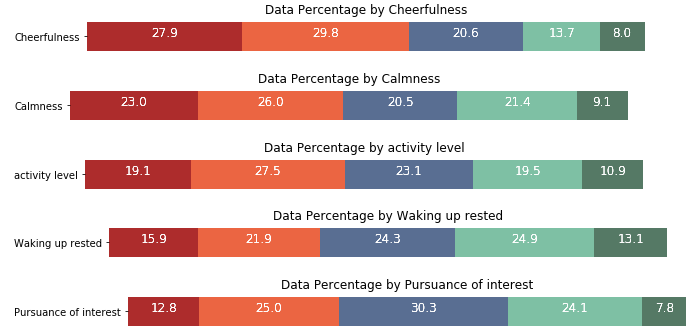}
    \caption{The change in emotional well-being during WFH}
    \label{fig:galaxy8}
\end{figure}

\item
\textbf{Proximity to the COVID-19 virus}: As exposure to the COVID-19 virus can affect the employees’ productivity, the employees were asked whether they, their family members, and roommates are affected with COVID-19 or not. They were also asked whether they were affected but got cured, or tested but didn’t get the results yet. Fig. 7 is the visual representation of the state of the employees' exposure to COVID-19. The findings are stated below.

\begin{enumerate}
    \item 
\textbf{Being affected with Covid-19}: From the data, it has been seen that 75\% of the respondents are not currently infected with COVID-19, were never infected and never even tested for the virus. As a result, their productivity is not hampered. 

\item
\textbf{Roommate infected with COVID-19}: The data reveals that 72.5\% of the respondents' roommates are not currently infected, were never infected and never even tested for the virus. Their productivity is not hampered due to this factor.

\item
\textbf{Family member infected with COVID-19}: If any of the family members of an employee is infected with COVID-19, his/her emotional well-being gets hampered causing decrease in productivity. From the data, it is seen that 70.5\% of the respondents’ family members are not currently infected, were never infected and never even tested for the virus. As a result, their productivity is not hampered due to this factor either.

\end{enumerate}

\item
\textbf{Miscellaneous}:
The respondents were asked about the change in other factors associated with productivity. The factors are the frequency of short breaks, freedom of time management decisions, frequency of listening music, frequency of household chores, and fear of losing job, as those factors impact productivity \cite{graziotin2014software} \cite{murphy2019predicts} \cite{kuvalekar2020job}. Fig. 8 is the visual representation of the change in these factors during WFH. The findings are stated below.

\begin{enumerate}
    \item
\textbf{The frequency of short breaks}: From related studies, it is seen that, short and frequent breaks increases productivity \cite{graziotin2014software}. In this survey, 68.9\% of the respondents said that the frequency of taking short breaks has increased during WFH or remained unchanged, which has a positive impact on their productivity.

\item
\textbf{Freedom of time management decisions}: According to literature, freedom of making decisions about managing someone's own time is positively associated with productivity \cite{murphy2019predicts}. From the survey data, it is clear that 78.7\% of the respondents’ freedom of time management decisions has increased or remained unchanged during WFH which has a positive impact on productivity.

\item
\textbf{Frequency of listening music}: Related works show that listening to music increases productivity \cite{graziotin2014software}. 63\% of the respondents of the survey said their frequency of listening to music has remained the same as before or increased, increasing productivity.

\item
\textbf{Frequency of household chores}: The amount of household chores has a significant impact on productivity. If the frequency of household chores increases, productivity decreases. From the data, it is seen that 65.2\% of the total respondents’ frequency of household chores increased during WFH, which has a negative impact on the productivity of the employees. 

\item
\textbf{Fear of losing job}: According to literature, job security is associated with productivity \cite{kuvalekar2020job}. The data reveals that 62.2\% of respondents said their fear of losing job has not increased, leaving no negative impact on their productivity.

\end{enumerate}

\begin{figure}[htp]
    \centering
    \includegraphics[width=8cm]{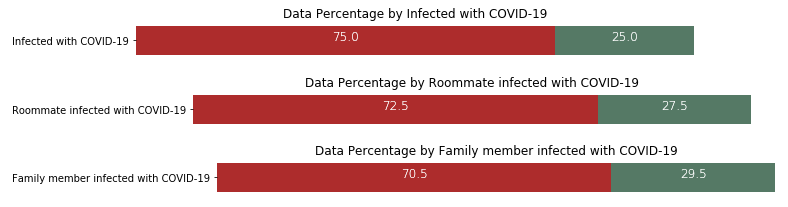}
    \caption{Proximity to COVID-19 during WFH}
    \label{fig:galaxy5}
\end{figure}

\begin{figure}[htp]
    \centering
    \includegraphics[width=8cm]{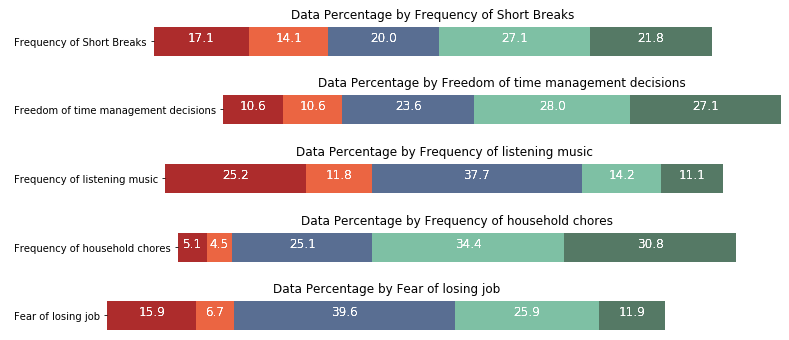}
    \caption{Miscellaneous Factors}
    \label{fig:galaxy6}
\end{figure}

\end{enumerate}

\section{Conclusion and Future Work}
Various studies have been done so far focusing on employee productivity. In this study, we focused on understanding employee productivity during WFH in Bangladesh to find out whether WFH can be established as an official system of operations in the IT sector of Bangladesh after the end of this lockdown through taking proper measures. After clearly defining the population for this research, factors related to their productivity were carefully chosen based on related studies. The questionnaire included only relevant questions. The pilot survey further helped in checking the relevance of the questionnaire. After incorporating feedback from the pilot survey, the questionnaire was modified accordingly. The randomly selected respondents of the final survey were asked how they perceive their current level of productivity and how the 24 aspects related to their work-life changed during WFH. Ensuring the anonymity of the respondents helped in getting accurate responses. Analyzing the responses and discovering the change in those 24 aspects, the longevity of WFH in the IT sector of Bangladesh has been discovered. Although the data reveal that WFH has a negative impact on 6 out of 24 of the productivity-related factors (the quality of internet connection, hardware accessibility, the frequency of household chores, workplace suitability, the frequency of interruptions, and cheerfulness of the employees), the drawbacks can be mitigated through proper steps. 

The outcomes of this study may be impacted by Hawthorne effect. That is, the participants’ responses might have been changed a bit because of knowing that they are participating in an experiment. Another issue to be considered is that, the emotional well-being related factors are not only affected by WFH. The panic of the pandemic might have further contribution to the employees’ state of emotional well-being.

There is a wide area of scope to proceed with the collected data and conduct further studies to get useful insights. Works in the future will include a detailed analysis of productivity, its correlation with the 24 aspects, and their correlation to each other. Answer for the questions like why WFH is affecting the mental well-being of the employees, how to deal with the work-space related issues and ensure a suitable work-environment in remote work, how to ensure better hardware access in WFH, how to better manage work-life and personal life to get the best out of WFH will be searched, which will require significant research efforts.

\medskip

\bibliographystyle{unsrt}
\bibliography{citations}

\end{document}